\input{aipcheck}

\documentclass[
    ,final            
  ]
  {aipproc}

\layoutstyle{6x9}

\begin{document}

\title{Top Physics at CDF}

\classification{14.65.Ha}
\keywords      {Tevatron, CDF, Standard Model, Top quark}

\author{Chang-Seong Moon\footnote{csmoon@fnal.gov, Speaker on behalf of the CDF collaboration}}{
  address={Department of Physics and Astronomy, Seoul National University,}\hspace{40mm} {Gwanakro Sillim-dong, Gwanak-gu, Seoul, 151-747 Korea}
}

\begin{abstract}
We present the recent results of top-quark physics using up to 6 fb$^{-1}$ of $p \bar p$ collisions 
at a center of mass energy of $\sqrt s$ = 1.96 TeV analyzed by the CDF collaboration. 
Thanks to this large data sample, precision top quark measurements are now a reality at the Tevatron. 
Further, several new physics signals could appear in this large dataset. 
We will present the latest measurements of top quark intrinsic properties as well as direct searches 
for new physics in the top sector. 

\end{abstract}

\maketitle

\section{Introduction}

Since the top quark has been discovered in 1995 by CDF~\cite{top_cdf} and D0~\cite{top_d0}
experiments at the Fermilab Tevatron collider,
the studies of top quark physics have been continued.
Up to now, the great performance of the Tevatron accelerator and the excellent operation of 
the CDF experiment allows to measure the top quark properties with improved precision 
and also explore several new physics signals in the top sector using large data samples.
The representative recent results on top quark physics from the CDF experiments with 6 fb$^{-1}$ 
are described due to the limited available space~\cite{cdf_top}.

\section{Top Physics}

The top quark which completes the third fermion generation in the Standard Model (SM), 
is the most massive of the known elementary particles.
This has led many physicists to believe that the top quark may shed light on the path to new physics.
Due to the large mass of the top quark, its properties allow predictions to be made of 
the mass of the Higgs boson under certain extensions of the SM.
At the Tavatron $p\bar p$ collider, top quark is mainly produced in pairs through quark and anti-quark (85\%)
or gluon-gluon fusion (15\%) and decays through the weak force almost exclusively into a $W$ boson and 
a $b$ quark.
The $t\bar t$ decays are labeled as dilepton, lepton+jets and all hadronic   
depending on whether a leptonic decay has occurred in both, one only or none of the two W bosons respectively.
The dilepton final state is only 5 percent of top pairs decay but
it is the easiest to identify because not very many other processes can produce such a striking final state.
The lepton+jets decays are considered as the golden channel
because of a sizable branching fraction combined with manageable background levels.
The all hadronic decay channel has the largest cross section 
but also a huge background from QCD multijet production.

\section{Top Pair Production Cross Section Measurement}

\begin{figure}[h]
    \centering
    \begin{minipage}[h]{.45\textwidth}    %
        {
      \hspace{1.5cm}
          \centering
          \includegraphics[width=0.72\textwidth, height=0.68\textwidth]{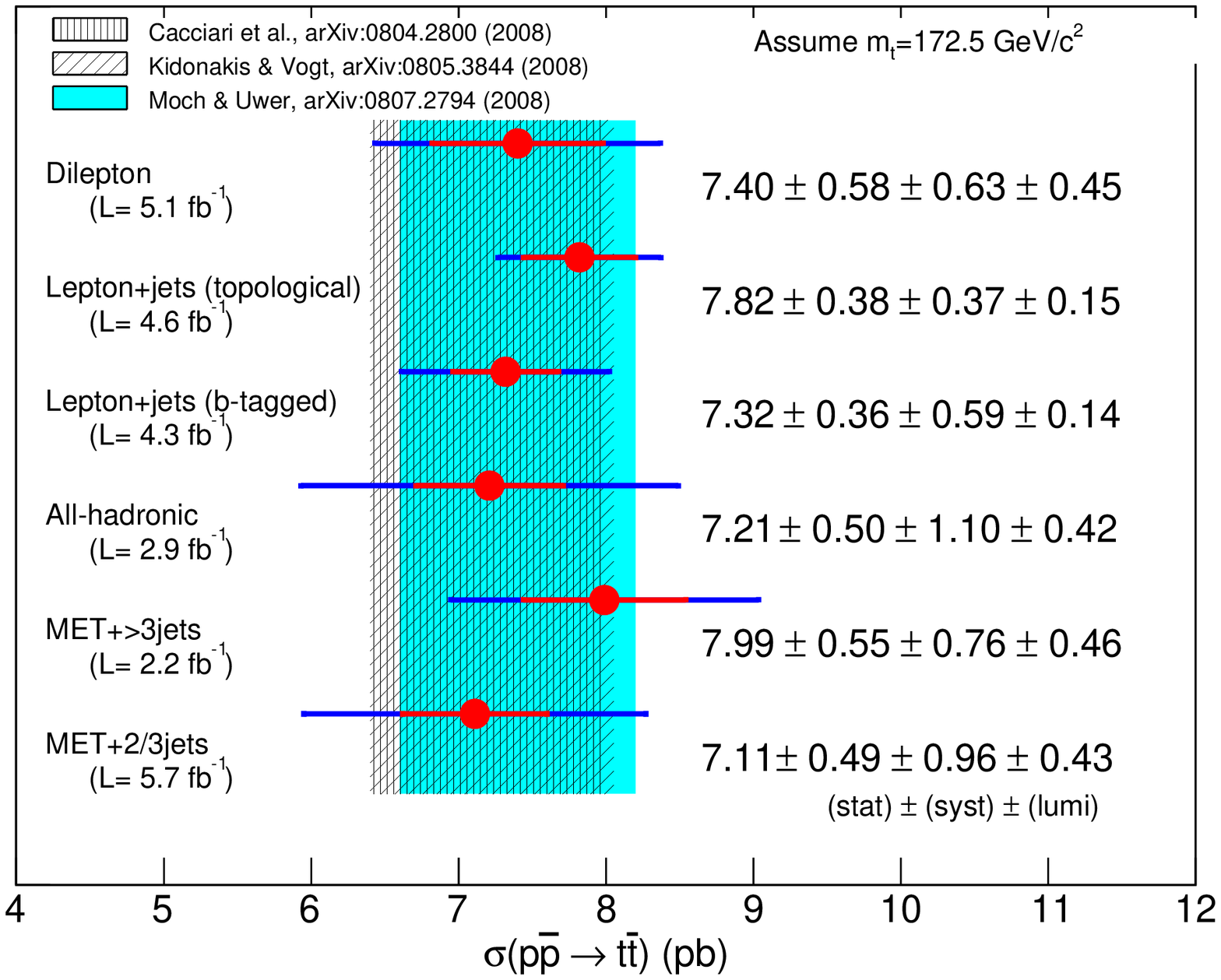}
        }
    \end{minipage}
    \hspace{0.7cm}
    \begin{minipage}[h]{.45\textwidth}    %
        \renewcommand{\arraystretch}{1.083}
        {
          \centering
          \includegraphics[height=0.68\textwidth]{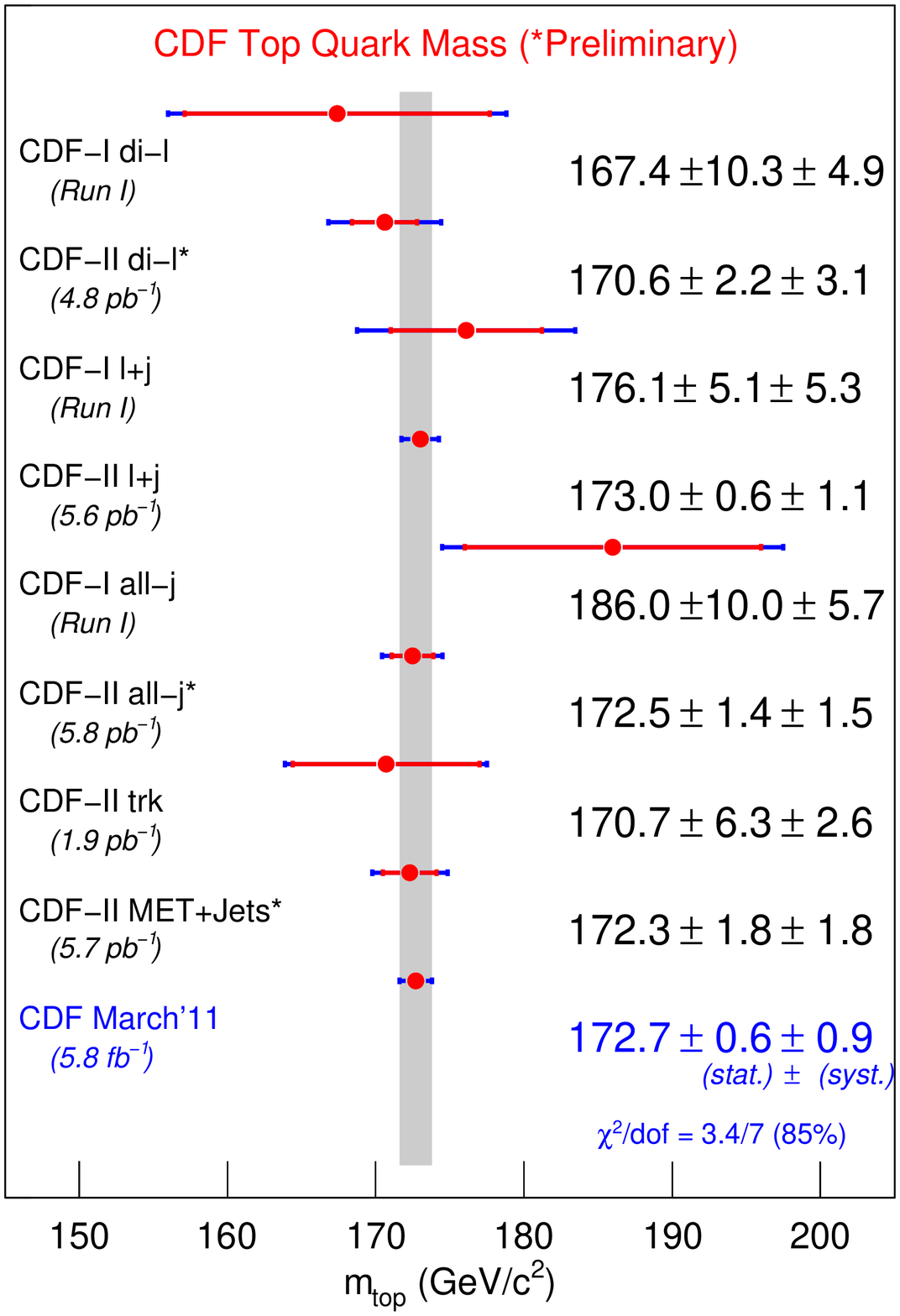}\\
        }
    \end{minipage}

  \caption{Left: Summary of CDF $t\bar t$ cross section measurements in different decay channels,
           Right: Summary of CDF top mass measurements in each decay channel}
  \label{plots_1}

\end{figure}

The SM predicts the $t\bar t$ production cross section at next-to-leading order (NLO) of $\sigma_{t\bar t}$
= 7.45$^{+0.72}_{-0.63}$ pb for the top quark mass of 172.5 GeV/$c^2$~\cite{xsec}.
A measurement of the $t\bar t$ production cross section provides a test of the QCD calculations 
and can probes new physics signals beyond the SM.
Most precise measurements come from the lepton+jets signature with two techniques~\cite{xsec_lj}.
First one used a topological selection based on a Neural Network (NN) exploiting the kinematical properties of 
the event and the other on $b$-tagging.
The latest analysis is performed in the dilepton signature~\cite{xsec_dil}.
In 5 fb$^{-1}$ of CDF data in the dilepton channel 343 signal candidates are found, the cross section 
is measured to be $\sigma_{t\bar t}$ = 7.4 $\pm$ 0.6 (stat) $\pm$ 0.6 (syst) $\pm$ 0.5 (lumi) pb.
After the requirement of at least one of the jets originating from a $b$ quark by the SecVtx algorithm, 
137 candidate events are found, for a measured cross section value of 
$\sigma_{t\bar t}$ = 7.3 $\pm$ 0.7 (stat) $\pm$ 0.5 (syst) $\pm$ 0.4 (lumi) pb.
CDF also measured the $\sigma_{t\bar t}$ with lower precision in all hadronic channel ~\cite{xsec_had}.
Figure~\ref{plots_1} (left) shows a summary of CDF $t\bar t$ cross section measurements which agree well
with the SM calculations.

\section{Measurement of Top Quark Mass}

The top quark mass is fundamentally important parameter of the Standard Model 
because of measurement of the top quark mass constraint the mass of the Higgs boson.
CDF present two new preliminary measurements using events in jets plus missing transverse energy 
and all hadronic final states channel using 6 fb$^{-1}$.  
The measured top masses are respectively $m_t$ = 172.3 $\pm$ 2.4 (stat+JES) $\pm$ 1.0 (syst) GeV/$c^2$
~\cite{mass_met}
and $m_t$ = 172.5 $\pm$ 1.4 (stat) $\pm$ 1.0 (JES) $\pm$ 1.2 (syst) GeV/$c^2$.~\cite{mass_had}
Figure~\ref{plots_1} (right) shows the CDF top mass measurements and the combination. 
The precision is reached to $\Delta M_{top}/M_{top}$ = 0.63\% and the uncertainty of top mass measurement
is only 1.09 GeV/$c^2$.
And the top mass different between top and anti-top quark is measured using 6 fb$^{-1}$ to be
$\delta$M$_{top}$ = -3.3 $\pm$ 1.7 
GeV/$c^2$ \cite{mass_diff}. It is consistent with CPT symmetry at a 2$\sigma$ level.

\begin{figure}[h]
    \centering
    \begin{minipage}[h]{.45\textwidth}    %
        {
          \centering
          \includegraphics[width=1.0\textwidth]{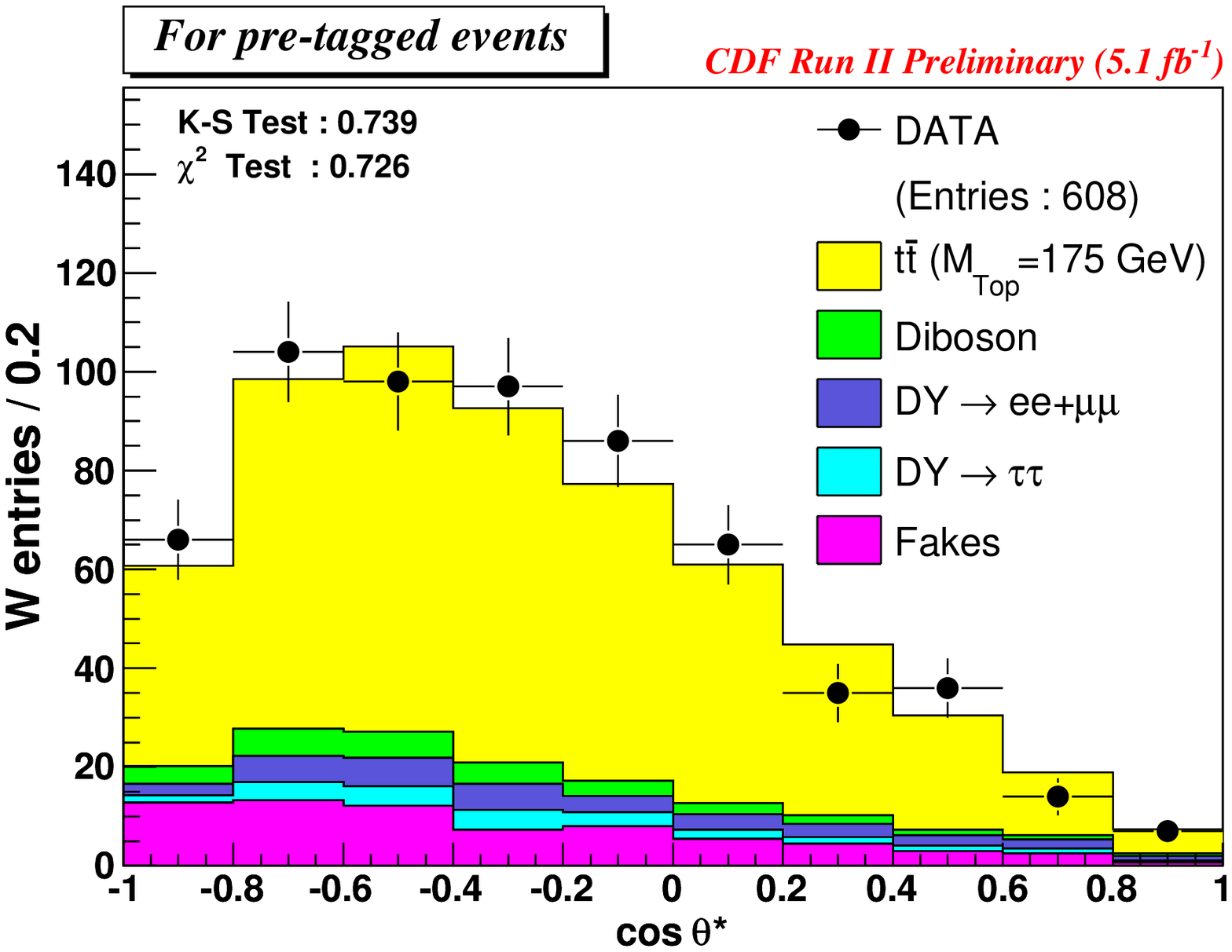}
        }
    \end{minipage}
    \hspace{0.5cm}
    \begin{minipage}[h]{.45\textwidth}    %
        \renewcommand{\arraystretch}{1.083}
        {
          \centering
          \includegraphics[height=0.72\textwidth]{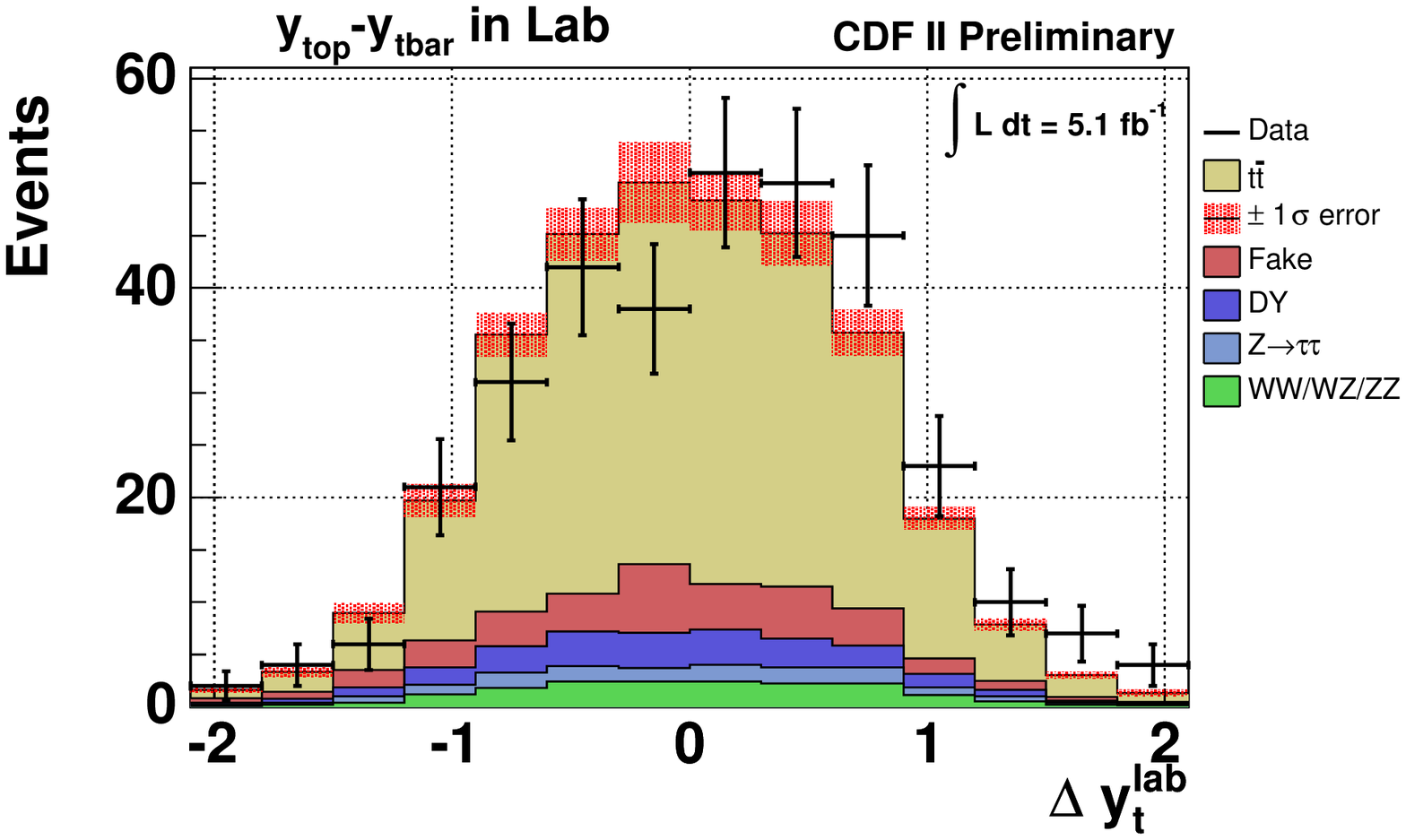}\\
        }
    \end{minipage}

  \caption{Left: The cos$\theta*$ distribution between data and the SM expectation in dilepton channel,
           Right: The $\Delta_y$ distribution of the top and anti-top in dilepton channel}
  \label{plots_2}

\end{figure}

\section{Study of Top Quark Properties}

Recent CDF results for the top quark properties are presented.
First, CDF have reported a measurement of the $W$ boson polarization in top quark decay. 
The $V$-$A$ structure of the weak interaction of the SM predicts that
the $W^{+}$ bosons from the top quark decay $t \rightarrow W^{+}b$ are
dominantly either longitudinally polarized ($\sim$70\%) or left-handed ($\sim$30\%),
while right-handed $W$ bosons are heavily suppressed and are forbidden
in the limit of massless b quarks.
The latest CDF measurements using 5 fb$^{-1}$ in dilepton channel present 
longitudinal fraction, $f_0 = 0.73 ^{+0.18}_{-0.17}$ (stat) $\pm$ 0.06 (syst)
and right-handed fraction, $f_{+}$ = -0.08 $\pm$ 0.09 (stat) $\pm$ 0.03 (syst).
Figure~\ref{plots_2} (left) shows the distribution of observable for $W$ boson polarization 
and good agreement between data and the SM predictions.

Next measurement is the forward-backward asymmetry, $A_{fb}$ of the top quark pair production.
The asymmetry can arise only in the next to leading order and is predicted to be about 0.078 
~\cite{afb}.
In the lepton+jets channel, the corrected parton level asymmetry is 0.158 $\pm$ 0.074. 
And the recent analysis using the dilepton events shows that at the parton level the measured 
$A_{fb}$ is 0.42 $\pm$ 0.15(stat) $\pm$ 0.05(syst) 
This result confirms the deviation from the theoretical predictions.
The rapidity difference between the top and anti-top quarks in the dileptonic decay mode is shown
in Figure~\ref{plots_2} (right).

\section{Searches in the Top Sector}

\begin{figure}[h]

    \centering
    \begin{minipage}[h]{.45\textwidth}    %
        {
          \hspace{0.7cm}
          \centering
          \includegraphics[height=0.62\textwidth, width=0.8\textwidth]{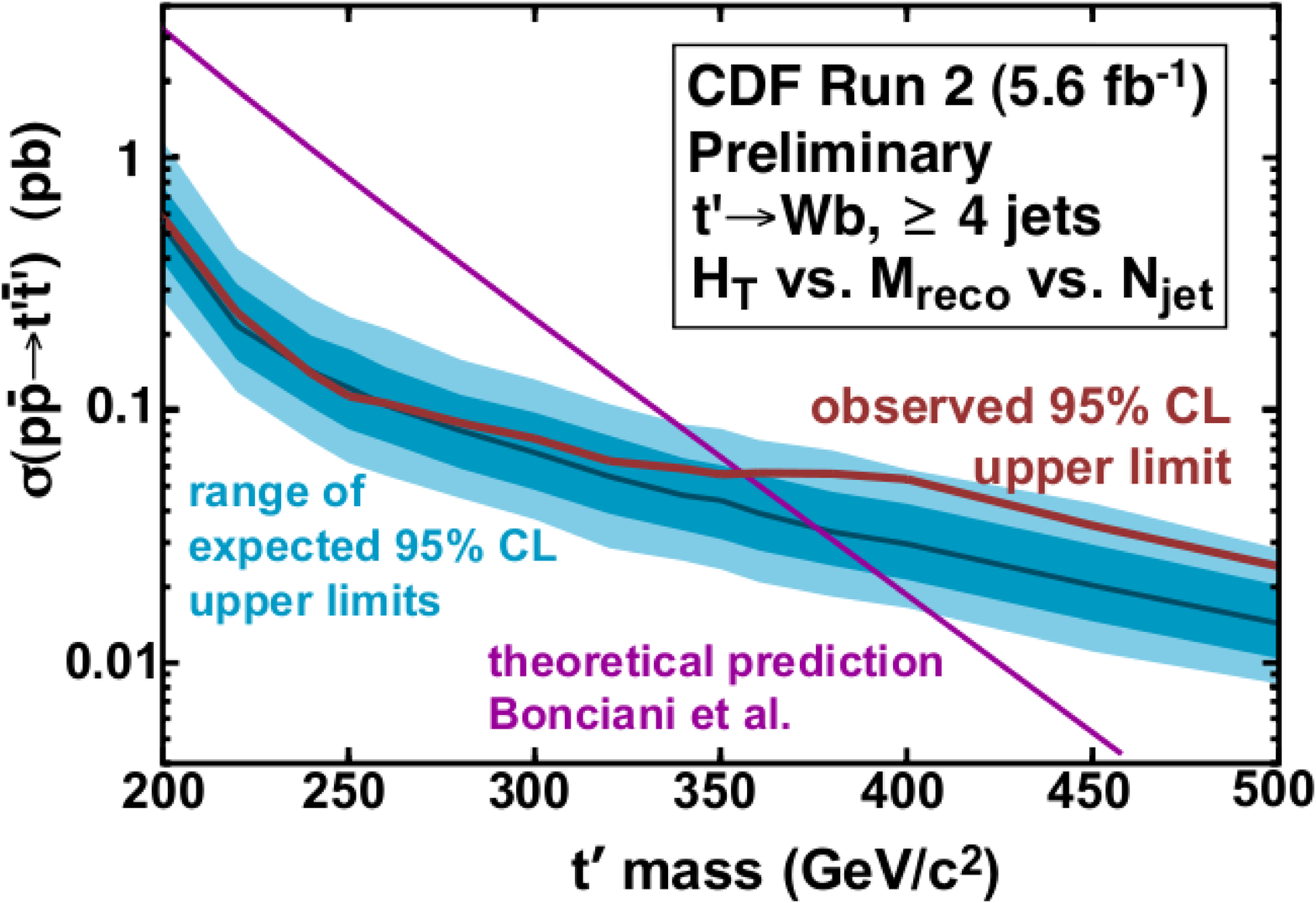}
        }
    \end{minipage}
    \hspace{0.5cm}
    \begin{minipage}[h]{.45\textwidth}    %
        \renewcommand{\arraystretch}{1.083}
        {
          \centering
          \includegraphics[height=0.62\textwidth]{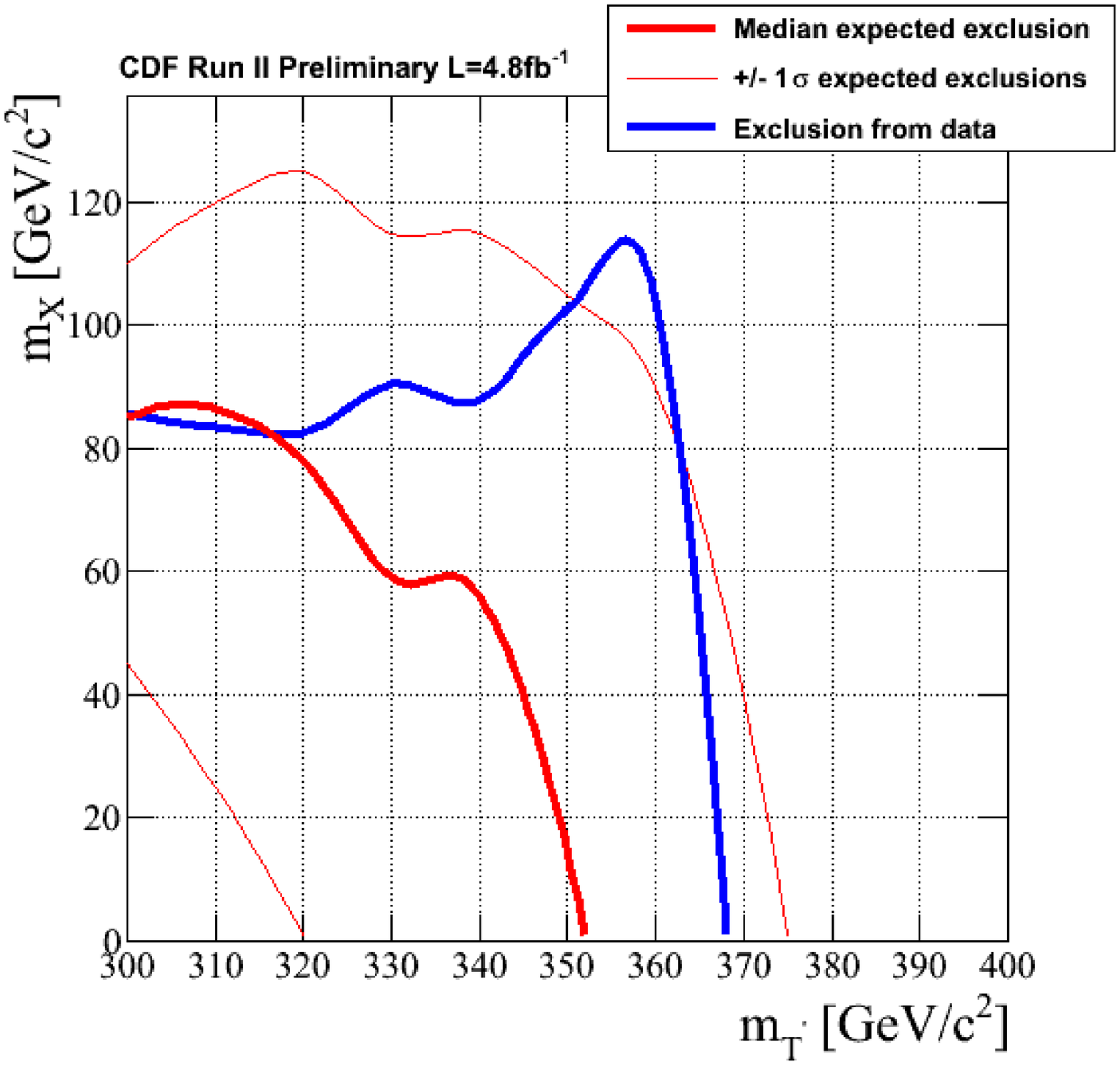}\\
        }
    \end{minipage}

  \caption{Left: Upper limit, at 95\% CL, on the production rate for $t^\prime$ as a function of $t^\prime$ mass (red),
           Right: Observed versus expected exclusion in ($m_{T^{\prime}}$, $m_X$)}
  \label{plots_3}

\end{figure}

CDF has searched for a heavy top quark $t^{\prime} \rightarrow Wb$ in the lepton+jets events using 6 fb$^{-1}$ 
~\cite{tprime1}. 
We exclude the SM fourth-generation $t^{\prime}$ quark with mass below 358 GeV/$c^2$ at 95\% C.L.
(see Figure~\ref{plots_3}(left)).
Another result is search for dark matter through the production of exotic $4^{th}$ generation 
quarks $t^{\prime}$ decaying via $t^{\prime} \rightarrow t + X$, where $X$ is dark matter
~\cite{tprime2}. 
We find that using 5 fb$^{-1}$ of data will allow for exclusion of 
$m_{T^{\prime}}$ < 360 GeV at 95\% C.L. for $m_{X}$ < 100 GeV. 
The Figure~\ref{plots_3} (right) describes the expected exclusion with $\pm 1\sigma$ vs observed exclusion.

\section{Conclusion}

We have presented recent results of top physics from CDF experiment
All the measurements are consistent with the SM prediction so far and new physics is not found yet in top sector.
More than 1000 reconstructed $t\bar t$ events in ~6 fb$^{-1}$ of dataset 
allow precision measurements of top quark properties. 
New results will be updated with 7$\sim$8 fb$^{-1}$ of data.
CDF top physics program and understanding of systematic effects 
will continue to play a significant role for years to come
using the expected more than 10 fb$^{-1}$ of CDF full dataset at the end of Run II of the Tevatron.

\begin{theacknowledgments}

I would like to thank my colleagues of the CDF collaboration for their hard work and dedication
to provide these rich physics results.
I also thank the organizers of DIS 2011 in Jefferson Lab for an interesting and successful conference. 

\end{theacknowledgments}

\bibliographystyle{aipproc}   

\end{document}